\documentstyle[12pt]{article}
\makeatletter
\@addtoreset{equation}{section}
\makeatother
\renewcommand{\theequation}{\thesection.\arabic{equation}}
\newcommand {\equ}[1] {(\ref{#1})}
\advance\hoffset by -7mm
\renewcommand{\title}[1]{\null\vspace{25mm}

\noindent{\Large{\bf #1}}\vspace{10mm}

\noindent {\large }}
\newcommand{\authors}[1]{\noindent{\large #1}\vspace{5mm}

}
\newcommand{\address}[1]{\noindent #1\vspace{10mm}

}
\renewcommand{\abstract}[1]{\vspace{17mm}

\noindent{\small{\em Abstract.} #1}\vspace{2mm}

}
\newcommand{\journal}[4]{{\em #1~}#2\,(19#3)\,#4;}
\newcommand{\aihp}{\journal {Ann. Inst. Henri Poincar\'e}}

\newcommand{\jmp}{\journal {J. Math. Phys.}}

\newcommand{\cmp}{\journal {Comm. Math. Phys.}}

\newcommand{\np}{\journal {Nucl. Phys.}}
\newcommand{\pl}{\journal {Phys. Lett.}}

\newcommand{\nc}{\journal {Nuovo Cim.}}

\newcommand{\annp}{\journal {Ann. Phys. (N.Y.)}}

\def\f{\frac}
\def\be{\begin{equation}}
\def\ee{\end{equation}}
\def\bea{\begin{eqnarray}}
\def\eea{\end{eqnarray}}
\def\bean{\begin{eqnarray*}}
\def\eean{\end{eqnarray*}}
\def\ba{\begin{array}} \def\ea{\end{array}}
\def\6{\partial}  
 \def\d{\delta}  \def\e{\epsilon}
 \def\h{\eta} 
  \def\l{\lambda}
 \def\n{\nu} \def\x{\xi}

  \def\O{\Omega}

\def\non{\nonumber\\}
\def\={\!\!\!&=&\!\!\!}
\def\n={\!\!\!&\not=&\!\!\!}
\def\+{\!\!\!&&\!\!\!+~}
\def\-{\!\!\!&&\!\!\!-~}

\begin{document}

\setcounter{page}{0}
\thispagestyle{empty}\hspace*{\fill} REF. TUW 94-19

\title{Algebraic structure of gravity in Ashtekar variables}

\authors{P. A. Blaga}
\address{Department of Geometry,
         University of Cluj (Romania)}
\authors{O. Moritsch\footnote{Work supported in part by the
         ``Fonds zur F\"orderung der Wissenschaftlichen Forschung''
         under Contract Grant Number P9116-PHY.},
         M. Schweda,
         T. Sommer\footnote{Work supported in part by the
         ``Fonds zur F\"orderung der Wissenschaftlichen Forschung''
         under Contract Grant Number P10268-PHY.},
         L. T\u{a}taru\footnote{Supported by the
         ``\"{O}sterreichisches Bundesministerium f\"{u}r
         Wissenschaft und  Forschung''.}
         \footnote{Permanent address Dept. Theor. Phys.,
         University of Cluj, Romania.}
         and H. Zerrouki\footnote{Work supported in part by the
         ``Fonds zur F\"orderung der Wissenschaftlichen Forschung''
         under Contract Grant Number P10268-PHY.}}
\address{Institut f\"ur Theoretische Physik,
         Technische Universit\"at Wien\\
         Wiedner Hauptstra\ss e 8-10, A-1040 Wien (Austria)}
\leftline{August 1994}
\abstract{
The BRST transformations for gravity in Ashtekar variables are
obtained by using the Maurer-Cartan horizontality conditions.
The BRST cohomology in Ashtekar variables is calculated with the
help of an operator $\delta$ introduced by S.P. Sorella \cite{sor},
which allows to decompose the exterior derivative as a BRST
commutator. This BRST cohomology leads to the differential
invariants for four-dimensional manifolds.
}


\newpage

\section{Introduction}

A great amount of work has been done recently in reformulation of
general relativity in terms of a new set of variables that replace
the spacetime metric \cite{ash}. These new variables, called Ashtekar
variables, have been intensively used in a large number of problems
in gravitational physics. Ashtekar's main task was the quantum
gravity issue and these new variables, indeed, have opened a novel
line of approach to it.

In any quantum field theory, one of the most important question is
the existence and the form of the anomalies. The anomalies,  as well
as the Schwinger terms and the invariant Lagrangians could be
calculated in a purely algebraic way by solving the Wess-Zumino
consistency condition \cite{wz}, which is equivalent to a tower of
descent equations, involving the nilpotent BRST operator $s$, as well
as the exterior spacetime derivative $d$.

The usual procedure for solving the descent equations is based on the
Russian formula and the transgression equation \cite{tm1} (see also
\cite{sto,witten,ba1,dviolette,band,ginsparg,tonin1,stora,brandt}).
However, for the gravity in Ashtekar variables it is difficult to
write down these equations and it is necessary to follow a different
scheme.

First, for the Ashtekar variables, the BRST transformations
\cite{brs} are different
from those obtained in the Yang-Mills case. Besides, the Russian
formula, in the form given for the Yang-Mills case \cite{sto}, does
not hold and we have to find out a new method for obtaining a
generalization of it in this case.
On the other hand, for the Ashtekar variables, as well as for the
gravity with torsion \cite{mss} it is difficult to write down a
transgression equation and to obtain the anomalies, the Schwinger
terms and the invariant Lagrangians.
All these quantities are BRST-invariants modulo $d$-exact terms and
they can be obtained by using an operator $\delta$ which allows to
express the exterior derivative $d$ as a BRST commutator
\be
d=-[s,\delta] \ .
\label{sorella}
\ee
Once the decomposition (\ref{sorella}) has been found, successive
applications of the operator $\delta$ on a polynomial $Q$ which
is a nontrivial solution of the equation
\be
sQ=0 \ ,
\label{wz1}
\ee
give an explicit nontrivial member of the BRST cohomology group
modulo $d$-closed terms.

The solving of the equation (\ref{wz1}) is a problem of local BRST
cohomology instead of a modulo-$d$ one. Therefore we see that, due
to the operator $\delta$, the study of the cohomology of
$s$ modulo $d$ can be reduced to the study of the local cohomology
of $s$ which, in turn, could be analyzed by using the spectral
sequences method \cite{dix}.

In this paper we prove the decomposition (\ref{sorella}) for gravity
in Ashtekar variables and we show that the operator $\delta$ offers
a straightforward way of classifying the BRST cohomology group for
these variables. In this way, we can give a cohomological
interpretation of the cosmological constant, of the Ashtekar
Lagrangian, as well as of the gravitational Chern-Simons terms.

The BRST transformations of the Ashtekar variables will be obtained
by making use of the geometrical formalism introduced by L. Baulieu
and J. Thierry-Mieg \cite{ba1,bt,tm2,bb}. This formalism allows us to
obtain the BRST transformations from Maurer-Cartan horizontality
conditions and it turned out \cite{mss,mss2,msst} (see also
\cite{fb}) to be very useful in the case of gravity with torsion,
as well as the gravity with Ashtekar variables.
Moreover, it allows to formulate the diffeomorphism transformations
of the Ashtekar variables as {\em local translations} in the tangent
space by means of the introduction
of the ghost field $\eta^a=\xi^{\mu}e^{a}_{\mu}$ where $\xi^{\mu}$
are the usual diffeomorphism ghosts and $e^a_{\mu}$ are the tetrads.
With these ghosts we can define the linear operator $\delta$ from the
decomposition (\ref{sorella}) in a very simple way.

The paper is organized as follows. In section 2, we recall the
Ashtekar variables, the Maurer-Cartan horizontality conditions and we
derive the BRST transformations and the Bianchi identities.
In section 3 we define the linear operator $\delta$ and we find
out a solution of the descent equations by using this operator and
the general method presented in Ref. \cite{st}.
Some explicit examples are presented in section 4 and a brief
discussion about the commutation properties of the $\delta$-operator
can be found in section 5. Finally, the appendices A and B are
devoted to some detailed calculations, respectively important
commutator relations in the tangent space and the determinant
of the tetrad.

\section{Ashtekar variables and the Maurer-Cartan horizontality
         conditions}

\setcounter{equation}{0}

\subsection{The Yang-Mills case}

In this case the fields are the Lie-valued 1-form gauge connection
$A=A^A_{\mu}T^A\/ dx^{\mu}$  and
the Lie-valued 0-form ghost field $c=c^AT^A$.  $T^A$ are the
antihermitian
generators of a finite representation  of the gauge group obeying
the following relations
$$
[T^A, T^B ]=f^{ABC}T^C \ ,
\hspace{2cm} Tr\{T^A T^B\}=\delta^{AB} \ .
$$

The BRST transformations
\bea
sA \= dc+Ac+cA=Dc \ , \nonumber\\
sc \= c^2 \ , \nonumber\\
s^2 \= 0 \ ,
\label{eq:ym}
\eea
could be obtained by reinterpreting (\ref{eq:ym}) as a Maurer-Cartan
horizontality condition (MCHC). In order to do it, we shall consider
the combined gauge-ghost field
\be
\tilde{A}=A+c
\label{eq:atil}
\ee
which could be considered as an Ehresmann connection on a principal
fibre bundle \cite{kn}, with the differential
\be
\tilde{d}=d-s \ ,
\hspace{1.5cm} \tilde{d}^2=0 \ .
\label{eq:dif}
\ee
The 2-form field strength $F$ is given by
\be
F=dA+A^2
\label{field}
\ee
and
\be
dF=[F,A]
\label{bi}
\ee
is its Bianchi identity. The field strength $\tilde{F}$ of the
connection $\tilde{A}$ is given by
\be
\tilde{F}=\tilde{d}\tilde{A}+\tilde{A}^2
\label{MC}
\ee
and it obeys the generalized Bianchi identity
\be
\tilde{d}\tilde{F}=[\tilde{F},\tilde{A}] \ .
\label{gbi}
\ee
The MCHC reads then
\be
\tilde{F}=F
\label{mchc}
\ee
and it splits into three components (\ref{eq:ym}) and
(\ref{field}) by expanding $\tilde{F}$ in terms of $A$ and $c$
and collecting the terms with the same form degree and ghost number.
Moreover, we have the Bianchi identity
\be
\tilde{d}\tilde{F}-[\tilde{F},\tilde{A}]=dF-[F,A]=0 \ .
\label{bi1}
\ee
It is important to emphasize that the BRST transformations
(\ref{eq:ym}) could be obtained directly from the
MCHC (\ref{mchc}).

\subsection{Ashtekar variables}

General relativity could be reformulated in term of two fields:
a {\em {real}} tetrad 1-form
\be
e^{a}=e^{a}_{\mu}dx^{\mu}
\label{tetrad}
\ee
and a {\em {complex self-dual}} connection 1-form
\be
A^{ab}=A^{ab}_{\mu}dx^{\mu} \ .
\label{conn}
\ee
Here the indices $(\mu,\nu,\dots)$ are spacetime indices
running from $0$ to $3$
while the indices $(a,b,c,\dots)$ are flat
tangent space indices  running also
from $0$ to $3$. The flat tangent space indices  are raised
and lowered with the Minkowski metric
\be
\eta^{ab}=\eta_{ab}=\left(
\begin{array}{cccc}
-1 & 0 & 0 & 0 \\
0 & 1 & 0 & 0 \\
0 & 0 & 1 & 0 \\
0 & 0 & 0 & 1
\end{array}
\right) \ .
\label{Min}
\ee
By definition, the complex Ashtekar
connection 1-form $A^{ab}$ is self-dual, i.e.
\be
*A^{ab}=i A^{ab} \ ,
\label{selfd}
\ee
where in the case of $SO(1,3)$ (in four dimensions)
\be
*A^{ab}=\frac{1}{2}\varepsilon^{ab}_{~~cd}A^{cd}
\ee
and $\varepsilon^{ab}_{~~cd}$ is the completely antisymmetric
tensor, with $\varepsilon^{0123}=1$ and $\varepsilon_{0123}=-1$.

The tetrad 1-form is related to the Ashtekar torsion 2-form by
\be
T^{a}=de^{a}+A^{a}_{~b}e^{b}=De^{a} \ ,
\label{tors}
\ee
where $D=d+A$ is the Ashtekar covariant exterior derivative.
This Ashtekar torsion 2-form does not vanish
even though the spin-connection $\omega^{ab}$, related to
$A^{ab}$ by the equation
\be
A^{ab}=\f1{2}( \omega^{ab}-i*\omega^{ab})=
\f1{2}( \omega^{ab}-\f{i}{2}\varepsilon^{ab}_{~~cd}\omega^{cd}) \ ,
\label{con2}
\ee
is torsion-free. The components of the
Ashtekar torsion, $T^{a}_{\mu\nu}$, are given by
\be
T^{a}_{\mu\nu}=\partial_{\mu}e^{a}_{\nu}-\partial_{\nu}e^{a}_{\mu}
+A^{a}_{~b\mu}e^{b}_{\nu} -A^{a}_{~b\nu}e^{b}_{\mu} \ .
\label{torc}
\ee
The  Ashtekar connection 1-form $A^{ab}$ is related to the
complex self-dual Ashtekar field strength 2-form by
\be
F^{ab}=dA^{ab}+A^{a}_{~c} A^{cb}= \frac{1}{2}F^{ab}_{\mu\nu}
dx^{\mu}dx^{\nu} \ ,
\label{YM}
\ee
where $F^{ab}$ is given by
\be
F^{ab}=\f{1}{2}(R^{ab}-i*R^{ab})
\label{ashcon}
\ee
with the curvature 2-form
\be
R^{ab}=d\omega^{ab}+\omega^a_{~c}\omega^{cb} \ .
\label{cur}
\ee

Applying the covariant exterior derivative to both sides of the
equations (\ref{tors}) and (\ref{YM}) one gets the Bianchi
identities
\bea
DT^{a} \= dT^{a}+A^{a}_{~b}T^{b}=F^{a}_{~b}e^{b} \ , \nonumber\\
DF^{ab} \= dF^{ab}+A^{a}_{~c}F^{cb}-F^a_{~c} A^{cb}=0 \ .
\label{Bi2}
\eea
It was proved \cite{rov} that the Ashtekar variables ($e^a, A^{ab}$)
are equivalent with the usual metric $g_{\mu\nu}$. This means that
if ($e^a, A^{ab}$) satisfy the equation of motion of the theory with
the action
\be
S=\int{F^{ab}\wedge e_a\wedge e_b} \ ,
\label{action}
\ee
then the metric
\be
g_{\mu\nu}(x)=e^a_{\mu}(x)e^b_{\nu}(x)\eta_{ab}
\label{metric}
\ee
is the solution of the Einstein equations. Viceversa, every solution
of the Einstein equations can be written in terms of the solutions
$(e^a, A_{ab})$  of the equations
of motion with the action (\ref{action}), as done in
eq.(\ref{metric}). In the case of Yang-Mills fields, the MCHC
implies the BRST transformations. Similiar conditions can be
formulated in the case of gravity \cite{ba1,mss,tm2,bb} and
it naturally includes the torsion.
Moreover, it allows to formulate the diffeomorphisms as
local translations. In this paper we shall write down the
MCHC for gravity in Ashtekar variables and we shall obtain
BRST transformations for the Ashtekar fields.

\subsection{Maurer-Cartan horizontality conditions}

The generalized tetrad-ghost field $\tilde{e}^{a}$ and the extended
complex self-dual connection-ghost field $\tilde{A}^{ab}$
are now defined as
\be
\tilde{e}^{a}=e^{a}+\eta^{a}
\label{tetr}
\ee
and
\be
\tilde{A}^{ab}=\hat{A}^{ab}+c^{ab} \ ,
\label{conA}
\ee
where $\eta^{a}$ is the ghost field of local translations in the
tangent space, $c^{ab}$ is the self-dual Ashtekar
ghost and $\hat{A}^{ab}$ is
given by
\be
\hat{A}^{ab}=A^{ab}_{~~c}\tilde{e}^{c}
=A^{ab}+A^{ab}_{~~c}\eta^{c} \ .
\label{ahat}
\ee
The 0-form $A^{ab}_{~~c}$ is defined by the expansion of the
0-form connection $A^{ab}_{\mu}$ in terms of the tetrad fields
$e^{a}_{\mu}$:
\be
A^{ab}_{~~\mu}=A^{ab}_{~~c}e^{c}_{\mu} \ .
\label{AA}
\ee
The ghost field of local translations $\eta^{a}$ is related to the
usual ghost for the diffeomorphisms $\xi^{\mu}$ by the relations
\bea
\eta^{a} \= \xi^{\mu}e^{a}_{\mu} \ ,
\nonumber\\
\xi^{\mu} \= E^{\mu}_{a}\eta^{a} \ ,
\label{ghosts}
\eea
where $E^{\mu}_a$ denotes the inverse of the tetrad $e^a_\mu$,
i.e.
\bea
e^{a}_{\mu} E^{\mu}_b \= \delta^a_b \ , \nonumber\\
e^{a}_{\mu} E^{\nu}_a \= \delta^{\nu}_{\mu} \ .
\eea
The generalized Ashtekar torsion 2-form and
Ashtekar field strength 2-form are given by
\bea
\tilde{T}^{a}\=\tilde{d}\tilde{e}^{a}+\tilde{A}^{a}_{~b}
\tilde{e}^{b} \ , \nonumber\\
\tilde{F}^{ab}\=\tilde{d}\tilde{A}^{ab}+\tilde{A}^{a}_{~c}
\tilde{A}^{cb} \ ,
\label{GTC}
\eea
and they obey the generalized Bianchi identities
\bea
\tilde{D}\tilde{T}^{a}\=\tilde{d}\tilde{T}^{a}
+\tilde{A}^{a}_{~b}\tilde{T}
^{b}= \tilde{F}^{a}_{~b}\tilde{e}^{b} \ ,
\nonumber\\
\tilde{D}\tilde{F}^{ab}\=
\tilde{d}\tilde{F}^{ab}+ \tilde{A}^{a}_{~c}\tilde{F}^{cb}-
\tilde{F}^{a}_{~c}\tilde{A}^{cb} =0  \ ,
\label{abi}
\eea
with $\tilde{D}$ the generalized covariant exterior
derivative.

Now we are able to formulate the Maurer-Cartan horizontality
conditions for the case of gravity in Ashtekar variables.
Following \cite{mss} we can say that these conditions state that
$\tilde{e}^{a}$ and all its generalized Ashtekar covariant exterior
differentials can be expanded over $\tilde{e}^{a}$
with classical (without tilde) coefficients, i.e.:
\be
\tilde{e}^{a}=\delta^{a}_{b}\tilde{e}^{b}\equiv horizontal \ ,
\label{hor1}
\ee
\be
\tilde{T}^{a}(\tilde{e},\tilde{A})=\frac{1}{2} T^{a}_{bc}(e,A)
\tilde{e}^{b}\tilde{e}^{c}\equiv horizontal \ ,
\label{hor2}
\ee
\be
\tilde{F}^{ab}(\tilde{A})=\frac{1}{2}F^{ab}_{~~cd}(A)
\tilde{e}^{c}\tilde{e}^{d}\equiv horizontal \ ,
\label{hor3}
\ee
where the 0-forms $T^a_{bc}$ and $F^{ab}_{~~cd}$ are
defined by the tetrad expansion of the Ashtekar 2-form
torsion (\ref{torc}) and the Ashtekar 2-form field
strength (\ref{YM}):
\be
T^{a}=\frac{1}{2}T^{a}_{bc}e^{b}e^{c} \ ,
\label{torc1}
\ee
\be
F^{ab}=\frac{1}{2}F^{ab}_{~~cd}e^{c}e^{d} \ .
\label{YM1}
\ee
It is worthwhile to remind that eq.(\ref{ahat}) is nothing
but the {\em {horizontality condition}} for the Ashtekar connection,
stating the fact that $\hat{A}^{ab}$ itself can be expanded over
$\tilde{e}$. The horizontality conditions (\ref{hor1})-(\ref{hor3})
are equivalent with the statements:
\bea
\tilde{e}^{a}\=\exp(i_{\xi})e^{a}=e^{a}+i_{\xi}e^{a} \ ,
\nonumber\\
\tilde{T}^{a}\=\exp(i_{\xi})T^{a}=
T^{a}+i_{\xi}T^{a}+ \frac{1}{2}i_{\xi}i_{\xi}T^{a} \ ,
\nonumber\\
\tilde{F}^{ab}\=\exp(i_{\xi})F^{ab}=F^{ab}+i_{\xi}F^{ab}+
\frac{1}{2}i_{\xi}i_{\xi}F^{ab} \ ,
\label{h123}
\eea
since $e^{a}$ is an 1-form, while $T^{a}$ and $F^{ab}$ are 2-forms.
These conditions reduce to the Russian formula when the
diffeomorphism transformation generated by $\xi$ is absent.

Now the MCHC for the case of gravity in Ashtekar variables
(\ref{hor1})-(\ref{hor3}) give, when expanded in terms
of the elementary fields ($e^{a},A^{ab},\eta^{a},c^{ab}$), the
nilpotent BRST transformations corresponding to the classical gauge
(Lorentz) rotations and to the diffeomorphism transformations.
The BRST transformations for the tetrad $e^{a}$ and for the
diffeomorphism ghost $\eta^{a}$ could be obtained
from (\ref{hor2}) which yields
\bea
de^{a}-se^{a}+d\eta^{a}-s\eta^{a}+\hat{A}^{a}_{~b}e^{b}+\hat{A}
^{a}_{~b}\eta^{b}+c^{a}_{~b}e^{b}+c^{a}_{~b}\eta^{b}= \non
=\frac{1}{2}T^{a}_{cd}e^{c}e^{d}+T^{a}_{cd}e^{c}\eta^{d}
+\frac{1}{2}T^{a}_{cd}\eta^{c}\eta^{d} \ .
\label{hor21}
\eea
Collecting the terms with the same ghost number
and form degree,  we can easily
obtain the BRST transformations for the  tetrad 1-form
$e^{a}$ and for the local
translation ghost $\eta^{a}$:
\bea
se^{a}\=d\eta^{a}+A^{a}_{~b}\eta^{b}+A^{a}_{~bc}\eta^{c}e^{b}
+c^{a}_{~b}e^b-T^{a}_{bc}e^{b}\eta^{c} \ , \non
\label{brt}
s\eta^{a}\=A^{a}_{~bc}\eta^{c}\eta^{b}+c^{a}_{~b}\eta^b
-\frac{1}{2}T^{a}_{bc}\eta^{b}\eta^{c} \ .
\label{brgd}
\eea

These equations could be rewritten in terms of the diffeomorphism
ghost $\xi^{\mu}$ which take the more familiar form
\cite{tm1,tm2}:
\bea
se^{a}_{\mu} \= c^{a}_{~b}e^{b}_{\mu}
-{\cal L}_{\xi}e^{a}_{\mu} \ , \nonumber\\
s\xi^{\mu} \= -\f1{2}{\cal L}_{\xi}\xi^{\mu} \ ,
\label{brtg}
\eea
where ${\cal L}_{\xi}$ denotes the Lie derivative~\cite{kn} along the
direction $\xi^\mu$, i.e.
$$
{\cal L}_{\xi}e^a_\mu=\xi^\lambda\partial_\lambda e^a_\mu+
(\partial_\mu\xi^\lambda)e^a_\lambda \ .
$$

\subsection{BRST transformations and Bianchi identities}

In this subsection we are going to give, for the convenience of
the reader, the BRST transformations and the Bianchi identities which
are contained in the Maurer-Cartan horizontality conditions
(\ref{hor1})-(\ref{hor3}) ~and from eqs.(\ref{GTC})
and (\ref{abi}) for each form sector and ghost number.

\begin{itemize}

\item {\bf Form sector two, ghost number zero $(T^{a}, F^{a}_{~b})$}

\bea
sT^{a}\=c^{a}_{~b}T^{b}+A^{a}_{~bk}\eta^{k}T^{b}
-F^{a}_{~b}\eta^{b}
\nonumber\\
\+A^{a}_{~b}T^{b}_{mn}e^{m}\h^{n}-F^{a}_{~bmn}e^{b}e^{m}\h^{n}
+(dT^{a}_{mn})e^{m}\h^{n}
\nonumber\\
\-T^{a}_{mn}e^{m}d\h^{n}+T^{a}_{mn}T^{m}\h^{n}
-T^{a}_{kn}A^{k}_{~m}e^{m}\h^{n} \ ,
\nonumber\\
sF^{a}_{~b}\=c^{a}_{~c}F^{c}_{~b}-c^{c}_{~b}F^{a}_{~c}
+A^{a}_{~ck}\h^{k}F^{c}_{~b}-A^{c}_{~bk}\h^{k}F^{a}_{~c}
\nonumber\\
\+A^{a}_{~c}F^{c}_{~bmn}e^{m}\h^{n}
-A^{c}_{~b}F^{a}_{~cmn}e^{m}\h^{n}+(dF^{a}_{~bmn})e^{m}\h^{n}
\nonumber\\
\+F^{a}_{~bmn}T^{m}\h^{n}
-F^{a}_{~bkn}A^{k}_{~m}e^{m}\h^{n}
-F^{a}_{~bmn}e^{m}d\h^{n} \ .
\label{FORMTWO}
\eea
For the Bianchi identities one has
\bea
DT^{a} \= dT^{a}+A^{a}_{~b}T^{b}=F^{a}_{~b}e^{b} \ , \nonumber\\
DF^{a}_{~b} \= dF^{a}_{~b}+A^{a}_{~c}F^{c}_{~b}
-A^{c}_{~b}F^{a}_{~c}=0 \ .
\label{BIG}
\eea

\item {\bf Form sector one, ghost number zero  $(e^{a}, A^{a}_{~b})$}

\bea
se^{a}\=d\h^{a}+A^{a}_{~b}\h^{b}+c^{a}_{~b}e^{b}
+A^{a}_{~bm}\h^{m}e^{b}
-T^{a}_{mn}e^{m}\h^{n} \ ,
\nonumber\\
sA^{a}_{~b}\=dc^{a}_{~b}+c^{a}_{~c}A^{c}_{~b}
+A^{a}_{~c}c^{c}_{~b}
+(dA^{a}_{~bm})\h^{m}+A^{a}_{~bm}d\h^{m}
\nonumber\\
\+A^{a}_{~c}A^{c}_{~bm}\h^{m}+A^{a}_{~cm}\h^{m}A^{c}_{~b}
-F^{a}_{~bmn}e^{m}\h^{n} \ .
\label{FORM1}
\eea

\item {\bf Form sector zero, ghost number zero
$( A^{a}_{~bm}, F^{a}_{~bmn}, T^{a}_{mn})$ }

\bea
sA^{a}_{~bm}\=-\h^{k}\6_{k}A^{a}_{~bm}-\6_{m}c^{a}_{~b}
+c^{a}_{~c}A^{c}_{~bm}-c^{c}_{~b}A^{a}_{~cm}
-c^{k}_{~m}A^{a}_{~bk} \ , \non
sT^{a}_{mn}\=-\h^{k}\6_{k}T^{a}_{mn}+c^{a}_{~k}T^{k}_{mn}
-c^{k}_{~m}T^{a}_{kn}-c^{k}_{~n}T^{a}_{mk} \ , \non
sF^{a}_{~bmn}\=-\h^{k}\6_{k}F^{a}_{~bmn}+c^{a}_{~c}F^{c}_{~bmn}
-c^{c}_{~b}F^{a}_{~cmn} \  \non
\-c^{k}_{~m}F^{a}_{~bkn}
-c^{k}_{~n}F^{a}_{~bmk} \ .
\eea
The Bianchi identities \equ{BIG} are projected on the
0-form Ashtekar torsion $T^a_{~mn}$ and on the
0-form Ashtekar field strength $F^a_{~bmn}$ to give:
\bea
dT^{a}_{mn}\=(\6_{k}T^{a}_{mn})e^{k} \non
\=(F^{a}_{~kmn}+F^{a}_{~mnk}+F^{a}_{~nkm} \non
\-A^{a}_{~bk}T^{b}_{mn}-A^{a}_{~bm}T^{b}_{nk}
-A^{a}_{~bn}T^{b}_{km} \non
\-T^{a}_{lk}T^{l}_{mn}-T^{a}_{lm}T^{l}_{nk}
-T^{a}_{ln}T^{l}_{km} \non
\+T^{a}_{lk}A^{l}_{~nm}+T^{a}_{ln}A^{l}_{~mk}
+T^{a}_{lm}A^{l}_{~kn} \non
\-T^{a}_{lk}A^{l}_{~mn}-T^{a}_{lm}A^{l}_{~nk}
-T^{a}_{ln}A^{l}_{~km} \non
\-\6_{m}T^{a}_{nk}-\6_{n}T^{a}_{km})e^{k} \ , \non
dF^{a}_{~bmn}\=(\6_{k}F^{a}_{~bmn})e^{k} \non
\=(-A^{a}_{~ck}F^{c}_{~bmn}-A^{a}_{~cm}F^{c}_{~bnk}
-A^{a}_{~cn}F^{c}_{~bkm} \non
\+A^{c}_{~bk}F^{a}_{~cmn}+A^{c}_{~bm}F^{a}_{~cnk}
+A^{c}_{~bn}F^{a}_{~ckm} \non
\-F^{a}_{~blk}T^{l}_{mn}-F^{a}_{~blm}T^{l}_{nk}
-F^{a}_{~bln}T^{l}_{km} \non
\+F^{a}_{~blk}A^{l}_{~nm}+F^{a}_{~bln}A^{l}_{~mk}
+F^{a}_{~blm}A^{l}_{~kn} \non
\-F^{a}_{~blk}A^{l}_{~mn}-F^{a}_{~blm}A^{l}_{~nk}
-F^{a}_{~bln}A^{l}_{~km} \non
\-\6_{m}F^{a}_{~bnk}-\6_{n}F^{a}_{~bkm})e^{k} \ .
\eea
We also have the additional equation
\bea
dA^{a}_{~bm}\=(\6_{n}A^{a}_{~bm})e^{n}\non
\=(-F^{a}_{~bmn}+A^{a}_{~cm}A^{c}_{~bn}-A^{a}_{~cn}A^{c}_{~bm}\non
\+A^{a}_{~bk}T^{k}_{mn}-A^{a}_{~bk}A^{k}_{~nm}
+A^{a}_{~bk}A^{k}_{~mn}+\6_{m}A^{a}_{~bn})e^{n} \ .
\eea

\item {\bf Form sector zero, ghost number one
           $(c^{a}_{~b}, \h^{a})$}

\bea
s\h^{a}\=c^{a}_{~b}\h^{b}+A^{a}_{~bm}\h^{m}\h^{b}
-\frac{1}{2}T^{a}_{mn}\h^{m}\h^{n} \ , \non
sc^{a}_{~b}\=c^{a}_{~c}c^{c}_{~b}
-\h^{k}\6_{k}c^{a}_{~b} \ .
\label{FORMZERO-GRAV}
\eea

\item {\bf Algebra between $s$ and $d$ }

{}From the above transformations it follows (see also appendix A):
\be
s^{2}=0 \ ,\hspace{1cm} d^{2}=0 \ ,
\ee
and
\be
\{s,d\}=0 \ .
\ee

\end{itemize}

\section{Solution of the descent equations}

\setcounter{equation}{0}
The question of finding the invariant Lagrangians,
the anomalies and the Schwinger terms for the four-dimensional
gravity in Ashtekar variables can be solved in a purely algebraic
way by solving the BRST consistency condition in the space of the
integrated local field polynomials.
In order to solve this question, we have to find out the
nontrivial solution of the equation
\be
s\Delta =0 \ ,
\label{e}
\ee
where $\Delta$ is an integrated local field polynomial, i.e.
$\Delta=\int{\cal A}$. The condition (\ref{e}) translates into the
local equation
\be
s{\cal A}+d{\cal Q}=0 \ ,
\label{sa}
\ee
where ${\cal Q}$ is some local polynomial and $d=dx^\mu \partial_\mu$
is the nilpotent exterior spacetime derivative which anticommutes
with the nilpotent BRST operator
$s$
\be
s^2=d^2=sd+ds=0 \ ,
\label{sd}
\ee
and it is {\em acyclic} (i.e. its cohomology group vanishes).

The local equation (\ref{sa}), due to the algebra
(\ref{sd}) and the acyclicity of $d$,
generates a tower of descent equations
\bea
s{\cal A}+d{\cal Q}^1=0 \nonumber\\
s{\cal Q}^1+d{\cal Q}^2=0 \nonumber\\
\cdots \nonumber\\
s{\cal Q}^{k-1}+d{\cal Q}^k=0 \nonumber\\
s{\cal Q}^k=0
\label{des.eq}
\eea
with ${\cal Q}^i$ local polynomials in the fields.

For the Yang-Mills case, these equations can be solved by means of a
transgression procedure generated by
the Russian formula (\ref{MC}) \cite{sto}.
More recently a new and efficient way of finding nontrivial
solutions of the tower (\ref{des.eq}) has been proposed by
S.P. Sorella \cite{sor} and successfully applied to the study of the
Yang-Mills cohomology \cite{st},
the gravitational anomalies \cite{ws}
and the algebraic structure of gravity with torsion \cite{mss,ms}.
The basic ingredient of the method is an operator
$\delta $ which allows us to express the exterior derivative $d$ as a
BRST commutator, i.e.:
\be
d=-[s,\delta] \ .
\label{comm}
\ee

Now it is easy to see that, once the decomposition~(\ref{comm}) has
been found, repeated application of the operator $\delta $ on the
polynomial ${\cal Q}$ which
is a nontrivial solution of  the last equation of (\ref{des.eq})
gives an explicit and nontrivial solution for the other cocycles
${\cal Q}^i$ and for ${\cal A}$.
If ${\cal A}$ has ghost number one then it is called an anomaly and
if it has ghost number zero then it represents an invariant
Lagrangian.
In other word using the operator $\delta$ we can calculate
the solution of the cohomology $H$($s$ mod $d$) if we know
the solution of the cohomology $H$($s$). Actually, as
has been shown in \cite{st}, the cocycles obtained by the descent
equations~(\ref{des.eq}) turn out to be completely equivalent to
those one based on the Russian formula.

For the gravity in Ashtekar variables the operator $\delta$
introduced in eq.(\ref{comm}) can be defined by
\bea
\delta\eta^a \= -e^a  \ , \nonumber\\
\delta \Phi \= 0
\hspace{0.8cm}
\mbox{for}
\hspace{0.8cm}
\Phi=(e^a,A^{ab},T^a,F^{ab},c^{ab}) \ .
\label{delta}
\eea
Now it is easy to verify that $\delta$ is of degree
zero\footnote{The degree is given by
the sum of the form degree and the ghost number.} and obeys the
following algebraic relations
\begin{equation}
\label{alg}d=- [s,\delta]
\hspace{0.7cm},
\hspace{0.7cm}
[d,\delta]=0 \ .
\end{equation}
In order to solve the tower (\ref{des.eq}) we shall make use of the
following identity
\be
e^{\delta}s=(s+d)e^{\delta} \ ,
\label{id}
\ee
which is a direct consequence of (\ref{alg}) (see \cite{st}).

Let us consider now the solution of the
descent equations (\ref{des.eq}) with a
given ghost number $G$ and form degree $N$,
i.e. a solution of the tower
\bea
s\Omega _4^G+d\Omega _3^{G+1}=0  \nonumber\\
s\Omega _3^{G+1}+d\Omega_2^{G+2}=0  \nonumber\\
s\Omega _2^{G+2}+d\Omega _1^{G+3}=0  \nonumber\\
s\Omega _1^{G+3}+d\Omega _0^{G+4}=0  \nonumber\\
s\Omega _0^{G+4}=0
\label{des1.eq}
\eea
with ($\Omega _4^G$, $\Omega _3^{G+1}$, $\Omega _2^{G+2}$,
$\Omega _1^{G+3}$, $\Omega_0^{G+4}$) local polynomials in the
variables ($e^a$, $A^{ab}$, $\eta^a$, $c^{ab}$) which,
without loss of generality, will be always
considered as irreducible elements, i.e. they cannot be expressed
as the product of several factored terms.
In particular $\Omega _4^0,\Omega _3^1$ and $\Omega _2^2$
correspond, respectively to an invariant Lagrangian, an anomaly
and a Schwinger term.

Due to the identity (\ref{id}) we can obtain the higher cocycles
$\Omega^{G+4-q}_q (q=1,2,3,4)$ once a nontrivial solution for
$\Omega^{G+4}_0$ is known.
Indeed, by applying the identity (\ref{id}) on $\Omega^{G+4}_0$
one gets
\be
(s+d)\left[e^\delta\Omega^{G+4}_0(\eta,c,A,T,F)\right]=0
\label{my}
\ee
But as one can see from eq.(\ref{delta}), the operator $\delta$ acts
as a translation on the ghost $\eta^a$ with an amount $(-e^a)$ and
eq.(\ref{my}) can be rewritten as
\be
(s+d)\Omega^{G+4}_0(\eta-e,c,A,T,F)=0 \ .
\label{my1}
\ee
Thus the expansion of the 0-form cocycle
$\Omega^{G+4}_0(\eta-e,c,A,T,F)$
in power of the 1-form tetrads $e^a$ yields all the cocycles
$\Omega^{G+4-q}_q$.

\section{Examples}

\setcounter{equation}{0}

In this section we want to apply the previous algebraic setup to
produce some interesting examples. We shall show that all interesting
objects which occur in Ashtekar theory as: the cosmological
constant, the Ashtekar action,
the action for the topological gravity and the Capovilla, Jacobson,
Dell action have a cohomological origin, i.e. they are solutions
of some descent equations. In the last step we investigate the
Chern-Simons terms in five dimensions. The examples are ordered by
the power of the Ashtekar field strength.

\subsection{The cosmological constant}

The simplest local BRST-invariant polynomial 0-form, which can be
defined, is given by
\be
\Omega^4_0(\eta)=\frac{1}{4!}\varepsilon_{abcd}\eta^{a}
\eta^{b}\eta^{c}\eta^{d} \ .
\label{cosm}
\ee
Since in four dimensions the product of five ghosts $\eta^a$
automatically vanishes it is easy to see that $\Omega^4_0$
represents a cohomology class of the BRST operator $s$, i.e.
\bea
s\Omega^4_0\=0~~~~~,~~~~~\Omega^4_0\not=s\widehat{\Omega}^3_0 \ .
\label{cosm1}
\eea
The corresponding 0-ghost term is the invariant Lagrangian
\be
\Omega^0_4=\f{1}{4!}\delta^4\Omega_0^4=
\f{1}{4!}\varepsilon_{abcd}e^a e^b e^c e^d=ed^4x \ ,
\ee
where $e$ is the determinant of the tetrad $e_{\mu}^a$ given in
the appendix B.

\subsection{Ashtekar Lagrangian}

This time we start with the cocycle
\be
\Omega^4_0=\f{1}{2i}\f{1}{2!}\varepsilon_{abcd}F^{ab}_{~~mn}
\eta^m\eta^n\eta^c\eta^d \ .
\label{ash1}
\ee
This cocycle is BRST-closed, $s\Omega^4_0=0$, but it is not
BRST-exact i.e.
$$
\Omega^4_0 \neq s\widehat{\Omega}^3_0 \ .
$$
For the case of  $SO(1,3)$ the invariant Lagrangian corresponding to
(\ref{ash1}) has the form
\bea
\Omega_4^0\=\f{1}{4!}\delta^4\Omega_0^4=\f{1}{4i}\varepsilon_{abcd}
F^{ab}_{~~mn}e^m e^n e^c e^d=F_{ab}e^ae^b \nonumber\\
\=iE^{\mu}_m E^{\nu}_n F^{mn}_{\mu\nu} e d^4x=i F^{mn}_{~~mn}e d^4x
\nonumber\\
\=\f1{2}e_{a\mu}e_{b\nu}F^{ab}_{~~\tau\sigma}\varepsilon^{\mu\nu
\tau\sigma}d^4x \ ,
\eea
where we have used the tetrad 1-forms $ e^a=e^a_{\mu}dx^{\mu} $,
 $e_a=\eta_{ab}e^b$ and the selfduality of the Ashtekar field
strength.
This is just the action introduced by Ashtekar \cite{ash}
(see also ~\cite{sp,hjm}), whose real part is the Palatini
Lagrangian.

\subsection{The topological action}

We also can build an action which is quadratic in the Ashtekar field
strength. In this case, using two Ashtekar field strengths one can
built up a BRST-invariant local polynomial
\be
\Omega_0^{4}=-\f{i}{2}\varepsilon_{abcd}F^{ab}_{~~kl}F^{cd}_{~~mn}
\eta^k\eta^l\eta^m\eta^n \ ,
\label{topgh}
\ee
to which it corresponds the invariant Lagrangian
\bea
\Omega_4^{0}\=-2 i\varepsilon_{abcd}F^{ab}F^{cd}
=4F^{ab}F_{ab} \nonumber\\
\=F^{ab}_{~~kl}F_{abmn}\varepsilon^{klmn}d^4x=F^{ab}_{~~\mu\nu}
F_{ab\tau\sigma}\varepsilon^{\mu\nu\tau\sigma} d^4x \ .
\eea

Witten  has suggested that
4D gravity has a phase described by a topological field
theory (TQFT) \cite{wit} (see also \cite{bbrt})
and in this phase the observables are global invariants.
In particular the Donaldson maps \cite{don} can be identified
as BRST-invariants of the corresponding TQFT. In order to extend
Witten's analysis for 4D usual gravity (not topological one) with
propagating degrees of freedom we have to describe these degrees
of freedom using variables that are related naturally to those
employed in TQFT. In fact
they must be suitable for implementing {\em both } diffeomorphism
and gauge invariance. The Ashtekar connection satisfies
these requirements since it replaces the metric and in the reduced
phase space, obtained after eliminating the constraints,
a restricted sector of the theory is described by the
Ashtekar-Renteln ansatz \cite{ar}
\be
F^{ab}=-\f{\lambda}{3}[e^ae^b-i*(e^ae^b)] \ .
\ee
With this ansatz the Ashtekar action becomes the topological action,
the BRST transformations of the Ashtekar connection coincides with
the corresponding one from TQFT \cite{cs1} if one identifies
$i_\xi F$ and $\f1{2}i_\x i_\x F$ with the usual ghosts
introduced in TQFT.

\subsection{An invariant Lagrangian with three Ashtekar fields}

In this case, using three Ashtekar field strenghts one can build up
the following 0-form $s$-cocycle
\be
\Omega_0^{4}=\f{1}{4}F^{ab}_{~~kl}F^{cd}_{~~mn}F_{ab}^{~~pq}
\varepsilon_{cdpq}\eta^k\eta^l\eta^m\eta^n \ .
\ee
Again this cocycle is not trivial, i.e. it cannot be
written as a $s$-coboundary
$$ \Omega_0^{4}\neq s\widehat\Omega_0^{3} \ . $$
This term leads to the following invariant Lagrangian
\be
\Omega_4^{0}=F^{ab}F^{cd}F_{ab}^{~~pq}\varepsilon_{cdpq} \ .
\ee
{}From this 4-form we can obtain the global invariant proposed by
Chang and Soo \cite{cs1,cs}.

\subsection{Capovilla, Jacobson, Dell Lagrangian}

In this case we shall try to build up a BRST cocycle with four
Ashtekar field strengths $F^{ab}$. Using the 0-form
Ashtekar field strength
$F^{ab}_{~~cd}$, one gets for the cocycle $\Omega_0^{4}$:
\be
\Omega_0^{4}=\f1{4}F^{ab}_{~~kl}F^{cd}_{~~mn}\varepsilon^{klmn}
F_{abpq}F_{cdrs}\eta^p\eta^q\eta^r\eta^s \ .
\label{cjdgh}
\ee
It can be easy checked that this $\Omega_0^{4} $ is $s$-closed,
i.e.
\be
s\Omega_0^{4}=0 \ .
\ee
Also here it can be identified with a cohomological class of the
BRST operator $s$ since
$$
\Omega_0^{4}\neq s \widehat{\Omega}_0^{3} \ .
$$
The cocycle (\ref{cjdgh}) gives rise to the invariant Lagrangian
\be
\Omega_4^{0}=F^{ab}_{~~kl}F^{cd}_{~~mn}
\varepsilon^{klmn}F_{ab}F_{cd} \ .
\label{cjd act}
\ee
Expression (\ref{cjd act}) is nothing but  the Capovilla,
Jacobson and Dell Lagrangian \cite{cjd} for the case of $SO(3,C)$.

It is interesting to remark that this action
depends only on the self-dual spin connection,
i.e. the Ashtekar variables, and a general scalar-density
Lagrange multiplier field, as a coefficient in front of
eq.(\ref{cjd act}). The spacetime metric does not appear in
this action in any form. On the other hand
Capovilla, Jacobson and Dell \cite{cjd}
have shown that the field equations, which follows
from this action, reproduce the Einstein equations and that
the spacetime metric can be built up entirely
from the Ashtekar field strength $F$. The metric in this case plays no
role whatsoever, although it can be reconstructed from the
Ashtekar field strength. They also showed that Ashtekar's formulation
of general relativity is contained in this new action.

\subsection{Chern-Simons term}

Let us discuss in details the construction of the
five dimensional Chern-Simons term. In this case the descent
equations take the form
\bea
&&s\O^{0}_{5}+d\O^{1}_{4}=0 \non
&&s\O^{1}_{4}+d\O^{2}_{3}=0 \non
&&s\O^{2}_{3}+d\O^{3}_{2}=0 \non
&&s\O^{3}_{2}+d\O^{4}_{1}=0 \non
&&s\O^{4}_{1}+d\O^{5}_{0}=0 \non
&&s\O^{5}_{0}=0
\eea
where, using Sorella's method \cite{sor}, the cocycles can be
obtained by
\bea
&&\O^{4}_{1}=\d\O^{5}_{0} \ ,\non
&&\O^{3}_{2}=\frac{\d^{2}}{2!}\O^{5}_{0} \ ,\non
&&\O^{2}_{3}=\frac{\d^{3}}{3!}\O^{5}_{0} \ ,\non
&&\O^{1}_{4}=\frac{\d^{4}}{4!}\O^{5}_{0} \ ,\non
&&\O^{0}_{5}=\frac{\d^{5}}{5!}\O^{5}_{0} \ .
\eea
In order to find a solution for $\O^{5}_{0}$ we use the redefined
Ashtekar ghost
\be
\hat{c}^{a}_{~b}=A^{a}_{~bm}\h^{m}+c^{a}_{~b} \ ,
\ee
which, from eq.(\ref{delta}), transforms as
\be
\d\hat{c}^{a}_{~b}=-A^{a}_{~b} \ .
\ee

For the 0-form cocycle $\O^{5}_{0}$ in five dimensions one gets
\bea
\O^{5}_{0}\=-\frac{1}{10}\hat{c}^{a}_{~b}
\hat{c}^{\hspace{0.04cm}b}_{~c}
\hat{c}^{\hspace{0.04cm}c}_{~d}
\hat{c}^{\hspace{0.04cm}d}_{~e}
\hat{c}^{\hspace{0.04cm}e}_{~a}
+\frac{1}{4}F^{a}_{~bmn}\h^{m}\h^{n}
\hat{c}^{\hspace{0.04cm}b}_{~c}
\hat{c}^{\hspace{0.04cm}c}_{~d}
\hat{c}^{\hspace{0.04cm}d}_{~a} \non
\-\frac{1}{4}F^{a}_{~bmn}\h^{m}\h^{n}F^{b}_{~ckl}\h^{k}\h^{l}
\hat{c}^{\hspace{0.04cm}c}_{~a} \ ,
\eea
which leads to the five dimensional Chern-Simons term
in Ashtekar variables
\be
\O^{0}_{5}= \frac{1}{10}A^{a}_{~b}A^{b}_{~c}A^{c}_{~d}
A^{d}_{~e}A^{e}_{~a}
-\frac{1}{2}F^{a}_{~b}A^{b}_{~c}A^{c}_{~d}A^{d}_{~a}
+F^{a}_{~b}F^{b}_{~c}A^{c}_{~a} \ .
\ee


\section{The ${\cal G}$-operator}
\setcounter{equation}{0}

In this section we want to compare the BRST structure
of the usual gravity with the gravity in Ashtekar
variables.
The BRST structure of gravity has been demonstrated
by Werneck de Oliveira and Sorella in \cite{ws}.
They showed that if we work in a local space generated by
the spin connection $\omega$, the curvature $R=d\omega+\omega^2$
and their ghosts then the exterior differential $d$
still does have the decomposition (\ref{comm}) but $d$
does not commute with $\delta$:
\be
2{\cal G}=[d,\delta]\neq 0 \ .
\ee
The ${\cal G}$-operator does not vanish even if we choose as
independent variables the Christoffel connecton
$\Gamma$, the Riemann tensor $R=d\Gamma+\Gamma^2$ and their ghosts
\cite{ws}.

However, if one uses tetrads $e^a_\mu$ together with the spin
connection $\omega$, as the independent variables in a first order
formalism developed by Palatini, then the operator
${\cal G}$ vanishes \cite{mss}. Moreover, using tetrads
for the Yang-Mills gauge fields in the presence of gravity
(with or without torsion) ${\cal G}$ vanishes
also \cite{mss},
in spite of the fact that for the pure YM case
one has ${\cal G}\neq0$.

So we can say that the diffeomorphisms carry, in some sense, the
action of the ${\cal G}$-operator through the tetrads.

\section{Conclusions}
\setcounter{equation}{0}

The present paper has shown that the algebraic structure of
gravity in the Ashtekar formalism could be entirely obtained from
the Maurer-Cartan horizontality conditions and by introducing an
operator $\delta$ which allows a useful decomposition of the
exterior spacetime differential as a BRST commutator.
This decomposition offers us a simple possibility to solve the
descent equations and to find some elements of the
BRST cohomology. In particular we have obtained the
actions for the free gravitational field proposed by Ashtekar
\cite{ash}
as well as Capovilla, Jacobson and Dell~\cite{cjd}.
The same technique can be applied to study  the
gravity in Ashtekar variables coupled with Yang-Mills
fields as well as to the characterization of the
Weyl anomalies in these variables \cite{msst}.

\section{Appendices:}

Appendix A is devoted to demonstrate the computation of some
commutators involving the tangent space derivative $\6_{a}$.
In appendix B one finds
some relations concerning the determinant of the tetrad
and the $\varepsilon$-tensor.


\section*{A~~~Commutator relations}

\setcounter{equation}{0}
\renewcommand{\theequation}{A.\arabic{equation}}

In order to find the commutator of two tangent space derivatives
$\6_{a}$, we make use of the fact that the usual spacetime
derivatives $\6_{\mu}$ have a vanishing commutator:
\be
[\6_{\mu},\6_{\nu}]=0 \ .
\ee
{}From
\be
\6_{\mu}=e^{m}_{\mu}\6_{m}
\ee
one gets
\bea
[\6_{\mu},\6_{\nu}]=0 \= [e^{m}_{\mu}\6_{m},e^{n}_{\nu}\6_{n}]\non
\= e^{m}_{\mu}e^{n}_{\nu}[\6_{m},\6_{n}]
+e^{m}_{\mu}(\6_{m}e^{n}_{\nu})\6_{n}
-e^{n}_{\nu}(\6_{n}e^{m}_{\mu})\6_{m}\non
\=e^{m}_{\mu}e^{n}_{\nu}[\6_{m},\6_{n}]
+(\6_{\mu}e^{k}_{\nu}-\6_{\nu}e^{k}_{\mu})\6_{k}\non
\=e^{m}_{\mu}e^{n}_{\nu}[\6_{m},\6_{n}]
+(T^{k}_{\mu\nu}-A^{k}_{~n\mu}e^{n}_{\nu}
+A^{k}_{~m\nu}e^{m}_{\mu}
)\6_{k}\non
\=e^{m}_{\mu}e^{n}_{\nu}(T^{k}_{mn}+A^{k}_{~mn}-A^{k}_{~nm}
)\6_{k} \non
\+e^{m}_{\mu}e^{n}_{\nu}[\6_{m},\6_{n}] \ ,
\eea
so that
\be
[\6_{m},\6_{n}]=-(T^{k}_{mn}+A^{k}_{~mn}-A^{k}_{~nm}
)\6_{k} \ .
\ee
\newline
For the commutator of $d$ and $\6_{m}$ we get
\bea
[d,\6_{m}]\=[e^{n}\6_{n},\6_{m}]\non
\=-(\6_{m}e^{k})\6_{k}-e^{n}[\6_{m},\6_{n}]\non
\=-(\6_{m}e^{k})\6_{k}+e^{n}(T^{k}_{mn}+A^{k}_{~mn}
-A^{k}_{~nm})\6_{k} \ ,
\eea
and one has therefore
\be
[d,\6_{m}]=(T^{k}_{mn}e^{n}+A^{k}_{~mn}e^{n}
-A^{k}_{~nm}e^{n} -(\6_{m}e^{k}))\6_{k} \ .
\ee
Analogously, from
\be
[s,\6_{\mu}]=0
\ee
one easily finds
\bea
[s,\6_{m}]\=(\6_{m}\h^{k}-c^{k}_{~m})\6_{k}
+\h^{n}[\6_{m},\6_{n}]\non
\=(\6_{m}\h^{k}-c^{k}_{~m}-T^{k}_{mn}\h^{n}
-A^{k}_{~mn}\h^{n}+A^{k}_{~nm}\h^{n})\6_{k} \ .
\eea


\section*{B~~~Determinant of the vielbein and
              the $\varepsilon$-tensor}

\setcounter{equation}{0}
\renewcommand{\theequation}{B.\arabic{equation}}

The definition of the determinant of the vielbein $e^{a}_{\mu}$
is given by
\bea
e\=det(e^{a}_{\mu})=\frac{1}{4!}\varepsilon_{a_{1}a_{2}a_{3}a_{4}}
\varepsilon^{\mu_{1}\mu_{2}\mu_{3}\mu_{4}}e^{a_{1}}_{\mu_{1}}
e^{a_{2}}_{\mu_{2}}
e^{a_{3}}_{\mu_{3}}e^{a_{N}}_{\mu_{N}} \ .
\eea
One can easily verify that the BRST transformation of $e$ reads
\be
se=-\6_{\l}(\x^{\l}e) \ .
\ee
For the case of $SO(1,3)$ one has
\bea
e^{0}e^{1}e^{2}e^{3}\=\frac{1}{4!}\e_{a_{1}a_{2}a_{3}a_{4}}
e^{a_{1}}e^{a_{2}}e^{a_{3}}e^{a_{4}}\non
\=\frac{1}{4!}\e_{a_{1}a_{2}a_{3}a_{4}}e^{a_{1}}_{\mu_{1}}
e^{a_{2}}_{\mu_{2}}e^{a_{3}}_{\mu_{3}}e^{a_{4}}_{\mu_{4}}
dx^{\mu_{1}}dx^{\mu_{2}}dx^{\mu_{3}}dx^{\mu_{4}}\non
\=\frac{1}{4!}\e_{a_{1}a_{2}a_{3}a_{4}}
\e^{\mu_{1}\mu_{2}\mu_{3}\mu_{4}}
e^{a_{1}}_{\mu_{1}}e^{a_{2}}_{\mu_{2}}e^{a_{2}}_{\mu_{2}}
e^{a_{4}}_{\mu_{4}}
dx^{0}dx^{1}dx^{2}dx^{3}\non
\=ed^{4}\!x=\sqrt{-g}d^{4}\!x \ ,
\eea
where $g$ denotes the determinant of the metric tensor $g_{\mu\nu}$
\be
g=det(g_{\mu\nu}) \ .
\ee
The $\varepsilon$-tensor has the usual norm
\be
\varepsilon_{a_{1}a_{2}a_{3}a_{4}}
\varepsilon^{a_{1}a_{2}a_{3}a_{4}}=-4! \ ,
\ee
and obeys the following relation under partial contraction
of two indices
\be
\varepsilon_{abcd}\varepsilon^{mncd}
=-2(\d^{m}_{a}\d^{n}_{b}-\d^{n}_{a}\d^{m}_{b}) \ ,
\ee
and in general the contraction of two $\varepsilon$-tensors is given
by the determinant of $\d$-tensors in the following way:
\be
\varepsilon_{a_{1}a_{2}a_{3}a_{4}}
\varepsilon^{b_{1}b_{2}b_{3}b_{4}}=-
\left|
\begin{array}{cccc}
\d^{b_{1}}_{a_{1}} & \d^{b_{2}}_{a_{1}}
&\d^{b_{3}}_{a_{1}}& \d^{b_{4}}_{a_{1}}  \\
\d^{b_{1}}_{a_{2}} & \d^{b_{2}}_{a_{2}}
&\d^{b_{3}}_{a_{2}}& \d^{b_{4}}_{a_{2}}  \\
\d^{b_{1}}_{a_{3}} & \d^{b_{2}}_{a_{3}}
&\d^{b_{3}}_{a_{3}}& \d^{b_{4}}_{a_{2}}  \\
\d^{b_{1}}_{a_{4}} & \d^{b_{2}}_{a_{4}}
&\d^{b_{3}}_{a_{4}}& \d^{b_{4}}_{a_{4}}
\end{array}
\right| \ .
\ee


{\bf ACKNOWLEDGEMENTS}

We are grateful to all the members of the Institut
f\"{u}r Theoretische Physik
of the Technische Universit\"{a}t Wien for
useful discussions and comments.
One of us (LT) would like to thank Prof. W. Kummer for the
extended hospitality at the institute.

\end{document}